\documentclass[aps,pra,showpacs,groupedaddress,onecolumn,showkeys]{revtex4-2}
\usepackage{amsmath}
\usepackage{graphicx}
\usepackage{diagbox}
\usepackage{xcolor}
\usepackage[colorlinks=true, letterpaper=true, pdfstartview=FitV, linkcolor=blue, citecolor=blue, urlcolor=blue]{hyperref}

\begin{document}

\title{Pure-quartic domain-wall solitons as topological bits for data transmission}

\author{Pengfei Li$^{1,2}$}
\email{lpf281888@gmail.com}
\author{Jun Ruan$^{1,2}$}
\author{Shilong Liu$^{3}$}
\author{Dumitru Mihalache$^{4}$}
\author{Boris A. Malomed$^{5,6}$}

\affiliation{$^{1}$Department of Physics, Taiyuan Normal University, Jinzhong, Shanxi 030619, China}
\affiliation{$^{2}$Institute of Computational and Applied Physics, Taiyuan Normal University, Jinzhong, Shanxi 030619, China}
\affiliation{$^{3}$femtoQ Lab, Department of Engineering Physics, Polytechnique Montréal, Montréal, QC H3T 1J4, Canada}
\affiliation{$^{4}$Horia Hulubei National Institute of Physics and Nuclear Engineering, Magurele, Bucharest RO-077125, Romania}
\affiliation{$^{5}$Department of Physical Electronics, School of Electrical Engineering, Faculty of Engineering, and Center for Light-Matter Interaction, Tel Aviv University, Tel Aviv 69978, Israel}
\affiliation{$^{6}$Instituto de Alta Investigaci\'{o}n, Universidad de Tarapac\'{a}, Casilla 7D, Arica, Chile}
\begin{abstract}
Domain walls (DWs) are topological defects produced by symmetry-breaking phase transitions. Although DWs have been the subject of much work due to their fundamental physical properties, they have not been explored in optical systems with higher-order dispersion. Recent experimental and theoretical works have demonstrated that pure-quartic (PQ) solitons, with their specific energy-width scaling, arise from the interplay of the quartic group-velocity dispersion (GVD) and Kerr nonlinearity. Here, we report solutions for PQ-DW solitons for the model of optical media with the PQ GVD. The analysis demonstrates that they are stable modes. Further investigation reveals their potential as data carriers for optical telecommunications. These results broaden the variety of optical solitons maintained by diverse nonlinear media.
\end{abstract}

\maketitle

\section{Introduction}
The interplay between the dispersion and nonlinearity plays a crucial role for the creation of conventional solitons \cite{RevModPhys.68.423}. The formation and stable propagation of these wave packets rely on the delicate
balance between the second-order group-velocity dispersion (GVD) and self-phase modulation (SPM) \cite{TURITSYN2012135}. Traditionally, studies of solitons were focused on the second-order (quadratic) GVD, while
higher-order GVD terms were often regarded as a detrimental perturbation, leading to unwanted emission of dispersive waves \cite{Hook:93}, soliton instabilities, and energy dissipation \cite{Kodama:94}.

This paradigm had shifted in 2016, when the experimental realization of pure-quartic (PQ) solitons was carried out in photonic-crystal waveguides by means of precise GVD engineering \cite{blanco2016pure}. Subsequently, PQ solitons have been observed in mode-locked laser incorporating an intra-cavity spectral pulse shaper \cite{runge2020pure}. Within the same regime, experimental studies have further extended this concept, demonstrating not only pure-quartic solitons but also higher-order species, including pure-sextic, -octic, and -decic solitons \cite{PhysRevResearch.3.013166}.

PQ solitons, emerging from the balance between SPM and fourth-order GVD, differ from the conventional ones by featuring oscillations in their exponentially decaying tails \cite{Tam:19}. The PQ soliton energy is inversely proportional to the third power of the pulse's temporal width, which implies that the single pulse energy of PQ soliton may be significantly higher than the energy of the conventional solitons with the same width. Another difference is that the fourth-order GVD destroys the system's Galilean invariance, making the creation of moving solitons a nontrivial issue \cite{PhysRevA.104.043526}. PQ solitons open an avenue for the development of a new branch of nonlinear optics and may lead to novel applications \cite{10.1063/5.0059525,luo2023research}. Various species of PQ solitons have been predicted in the framework of the nonlinear Schr\"{o}dinger equation (NLSE) with the dominant fourth-order GVD, including dissipative \cite{Taheri:19,Qian:22,Wu:24,Wu:25}, Raman \cite{Liu:21,Wang:22}, dark \cite{Alexander:22,LI2022111950,Parra-Rivas:22,Gao:24,PhysRevA.111.013510}, asymmetric \cite{Li:24}, spatiotemporal \cite{PhysRevA.109.033516} PQ solitons, and their bound states (\textquotedblleft molecules") \cite{Deng:25}, as well as the pure-high-even-order dispersion bound solitons \cite{han2024pure}. It has been found that PQ solitons are lowest-order members of an infinite hierarchy of soliton modes that arise from the interplay of SPM and higher-even-order GVD, if it appears as the single dispersive term in the model \cite{PhysRevResearch.3.013166,han2024pure}. More general soliton modes have been reported in the framework of the generalized NLSE including both quadratic and quartic GVD terms \cite{PhysRevA.101.043822,Zhang:24,Liu:24,PhysRevA.109.053528,Zou:25,Wang:25,ZHANG2025116094}.

The subject of the present work are solitons of the domain-wall (DW) type, produced by a pair of coupled NLSEs with the fourth-order GVD terms. This species of soliton modes was not considered previously. The DW is a topological defect connecting two half-infinite domains filled by different mutually immiscible stable states of a two-component physical system. In addition to commonly known DWs in magnetism \cite{landau2013electrodynamics}, a relevant example is DWs in thermal convection \cite{Steinberg_1985}, modeled by a pair of nonlinearly coupled Ginzburg-Landau equations for amplitudes of two roll patterns with different orientations \cite{PhysRevA.42.7244,MALOMED2022127802}. Domain-wall solitons, i.e., structures separating adjacent domains carrying immiscible field states, are known in nonlinear optics \cite{HAELTERMAN1994265,PhysRevE.50.1565,Haelterman:94,haelterman,Sheppard:94,PhysRevLett.81.1409,Zhang:10,Zhang:11,Kartashov:12,PhysRevA.89.063812,ahmad2016domain,Wang:17,gilles2017polarization,10.1063/1.5091811}, and binary Bose-Einstein condensates \cite{PhysRevLett.87.140401,PhysRevA.65.063621,PhysRevE.106.054207}. They are modeled by systems of coupled NLSEs. Our objective is to reveal DWs produced by NLSE systems including the pure-quartic GVD, SPM, and XPM (cross-phase modulation). In particular, our study shows that they may be used as topological bits for optical data-encoding transmission.

The starting point is the propagation of the two polarization components of the optical field in a medium with higher-even-order GVD and cubic self-focusing or defocusing nonlinearity, which is represented, respectively, by coefficient $\gamma >0$ or $\gamma <0$. This setting is described by the system of one-dimensional modified NLSEs,
\begin{equation}
i\frac{\partial A_{1}}{\partial Z}=-i^{m}\frac{\beta _{m}}{m!}\frac{\partial^{m}A_{1}}{\partial T^{m}}-\gamma (|A_{1}|^{2}+2|A_{2}|^{2})A_{1},
\label{NLSE1}
\end{equation}%
\begin{equation}
i\frac{\partial A_{2}}{\partial Z}=-i^{m}\frac{\beta _{m}}{m!}\frac{\partial^{m}A_{2}}{\partial T^{m}}-\gamma (|A_{2}|^{2}+2|A_{1}|^{2})A_{2},
\label{NLSE2}
\end{equation}%
where $A_{1}\left( Z,T\right) $ and $A_{2}\left( Z,T\right) $ are slowly varying amplitude of the two components, $Z$ is the propagation distance, $T$ is the retarded time, $T=t-\beta _{1}Z$, in the reference frame moving with the group velocity of the carrier wave, and even number $m$ is the GVD order, with the respective coefficient $\beta _{m}$.

\begin{figure}[th]
	\centering
	\includegraphics[width=0.5\linewidth]{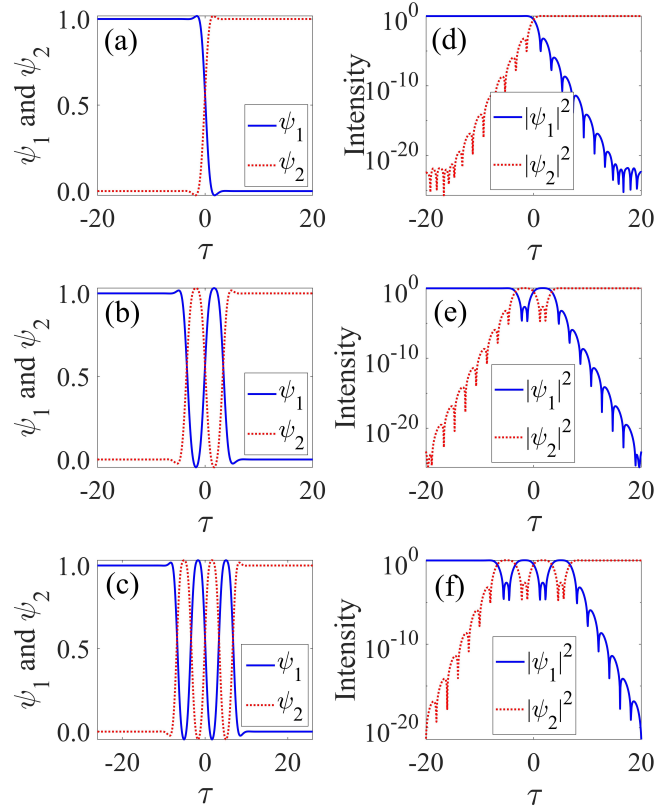}
	\caption{PQ-DW solitons solutions produced by the numerical solution of Eqs. (\protect\ref{system_psi1}) and (\protect\ref{system_psi2}) with propagation constant $-\protect\beta =-1$ for different widths $\protect\tau_{0}$ in input (\protect\ref{ansatz}): (a,d) $\protect\tau _{0}=2$; (b,e) $\protect\tau _{0}=6$; (c,f) $\protect\tau _{0}=12$. The left column displays the temporal profiles on the linear scale, while the right column shows the corresponding intensities ($|\psi_{1}|^{2}$ and $|\psi_{2}|^{2}$) on the log scale.}
	\label{fig1}
\end{figure}

Here, we consider pure quartic GVD with $m=4$ in Eqs. (\ref{NLSE1}) and (\ref{NLSE2}), where $\beta _{4}<0$ and $\beta _{4}>0$ stand for the anomalous and normal quartic GVD, respectively. PQ-DW solutions exist under condition $\beta _{4}\gamma >0$, hence for the anomalous quartic GVD ($\beta _{4}<0$), the PQ-DW solutions can exist if $\gamma <0$ (self-defocusing). By means of rescaling, $A_{1}\left( Z,T\right) =\sqrt{P}/N\Psi _{1}\left( z,\tau \right)$, $A_{2}\left( Z,T\right) =\sqrt{P}/N\Psi _{2}\left( z,\tau \right) $, Eqs. (\ref{NLSE1}) and (\ref{NLSE2}) are cast in the dimensionless form:
\begin{equation}
i\frac{\partial \Psi _{1}}{\partial z}-\frac{1}{24}\frac{\partial ^{4}}{\partial \tau ^{4}}\Psi _{1}-(|\Psi _{1}|^{2}+2|\Psi _{2}|^{2})\Psi _{1}=0,
\label{NLSE3}
\end{equation}
\begin{equation}
i\frac{\partial \Psi _{2}}{\partial z}-\frac{1}{24}\frac{\partial ^{4}}{\partial \tau ^{4}}\Psi _{2}-(|\Psi _{2}|^{2}+2|\Psi _{1}|^{2})\Psi _{2}=0,
\label{NLSE4}
\end{equation}
where $z=Z/L_{D}$ and $\tau =T/T_{0}$ are the normalized propagation distance and time, with GVD length $L_{D}=T_{0}^{4}/\left\vert \beta_{4}\right\vert $ corresponding to an incident pulse with temporal width $T_{0}$. Further, the pulse's peak power $P$ defines the nonlinearity length $L_{\mathrm{NL}}=1/\left( \left\vert \gamma \right\vert P\right) $ and the soliton number, $N=\sqrt{L_{D}/L_{\mathrm{NL}}}$. For $m>2$, Eqs. (\ref{NLSE1}) and (\ref{NLSE2}) have no known exact analytic solutions for solitary pulses, therefore, the following consideration is based on numerical methods. The corresponding physical system to perform these couple equations could be implemented using two optical waves propagating in a fiber-optic cavity, in which the fiber's nonlinearity supplies XPM and SPM, and a pulse shaper is employed to engineer the effective quartic dispersion. Such a configuration offers a promising platform for exploring advanced nonlinear pulse shaping regimes, including the PQ-DW solitons discussed below.

We look for stationary PQ-DW solutions to Eqs. (\ref{NLSE3}) and (\ref{NLSE4}) with the real temporal profiles $\psi _{1,2}$ and real propagation constant $-\beta $ as
\begin{equation}
\left\{ {\Psi _{1}\left( z,\tau \right) ,\Psi _{2}\left( z,\tau \right) }\right\} =e^{-i\beta z}\left\{ {\psi _{1}(\tau ),\psi _{2}(\tau )}\right\} .
\label{Psi1Psi2}
\end{equation}
The substitution of this ansatz in Eqs. (\ref{NLSE3}) and (\ref{NLSE4}) leads to a set of equations for the temporal profiles $\psi _{1,2}$:
\begin{equation}
-\frac{1}{24}\frac{d^{4}\psi _{1}}{d\tau ^{4}}-( \psi_{1} ^{2}+2 \psi _{2} ^{2})\psi _{1}+\beta\psi _{1}=0,
\label{system_psi1}
\end{equation}
\begin{equation}
-\frac{1}{24}\frac{d^{4}\psi _{2}}{d\tau ^{4}}-( \psi_{2} ^{2}+2 \psi _{1} ^{2})\psi _{2}+\beta\psi _{2}=0.
\label{system_psi2}
\end{equation}
We solved Eqs. (\ref{system_psi1}) and (\ref{system_psi2}) numerically, using the Newton-conjugate gradient method \cite{yang2010nonlinear}. With an appropriate initial guess, the method converges to numerically exact stationary solutions through successive corrections.

We sought PQ-DW solutions of Eqs. (\ref{system_psi1}) and (\ref{system_psi2}), starting from the natural DW-shaped initial guess,
\begin{equation}
\left\{ \psi _{1},\psi _{2}\right\} =\frac{\sqrt{\beta }}{2}\left[ 1-\tanh (\tau /\tau_{0}),1+\tanh (\tau /\tau _{0})\right],
\label{ansatz}
\end{equation}
where $\sqrt{\beta }$/2 and $\tau _{0}$ are the amplitude and width of the ansatz.\ The PQ-DW solution links two different continuous-wave (CW) states, $\left\{ \psi _{1},\psi _{2}\right\} =\sqrt{\beta }\left\{ 1,0\right\} $ and $\sqrt{\beta }\left\{ 0,1\right\} $, that fill the domains at $\tau\rightarrow \pm \infty $.

\begin{figure}[th]
\centering
\includegraphics[width=0.5\linewidth]{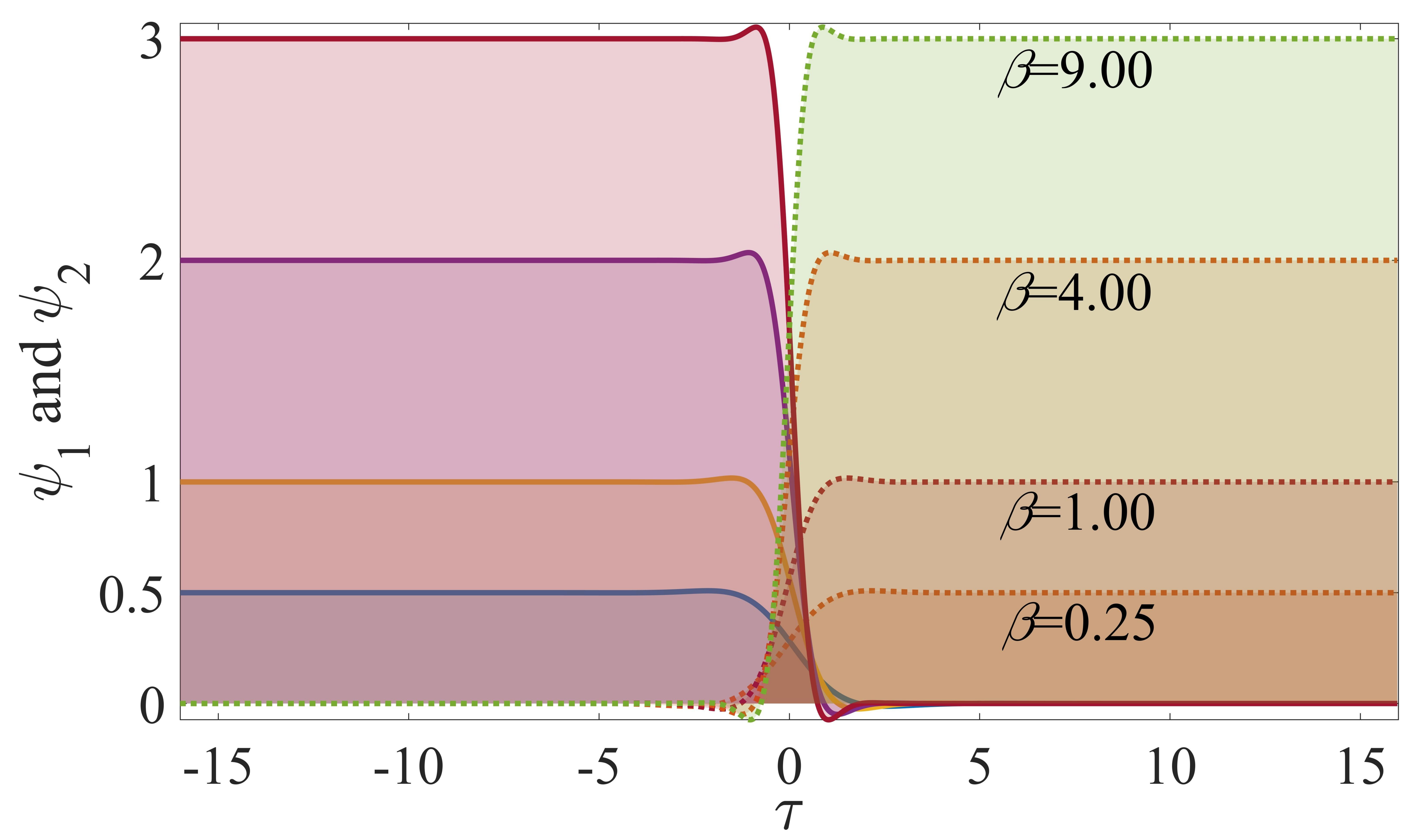}
\caption{Examples of single-kink PQ-DWS solitons with propagation constants $-\protect\beta =-0.25$,$-1$, $-4$, and $-9$, respectively.}
\label{fig2}
\end{figure}

Numerically calculated temporal and intensity profiles of the PQ-DW solutions for $\beta =1$ with different widths $\tau _{0}$ in input (\ref{ansatz}) are summarized in Fig. \ref{fig1}. At variance with the smooth hyperbolic-tangent-like shapes of conventional DWs in the system with the quadratic dispersion \cite{PhysRevE.106.054207}, these shapes feature oscillations in exponentially decaying tails, which is an effect of the fourth-order GVD \cite{Tam:19}. Panels \ref{fig1}(a) and \ref{fig1}(d)
represent the PQ-DW soliton with the single kink in each component, while panels \ref{fig1}(b) and \ref{fig1}(e) show kink-antikink-kink ($\psi _{1}$) and antikink-kink-antikink ($\psi _{2}$) complexes. More complex multikink states are seen in Figs. \ref{fig1}(c) and \ref{fig1}(f). We have checked that the PQ-DW solitons extend by including more kink-antikink pairs into the complexes with the increase of the input's width. While the above examples of single- and multi-kink solutions were produced for fixed $\beta=1$, a set of PQ-DW single-kink solitons obtained with different values of $\beta $ are presented in Fig. \ref{fig2}.

\begin{figure}[th]
\centering
\includegraphics[width=0.5\linewidth]{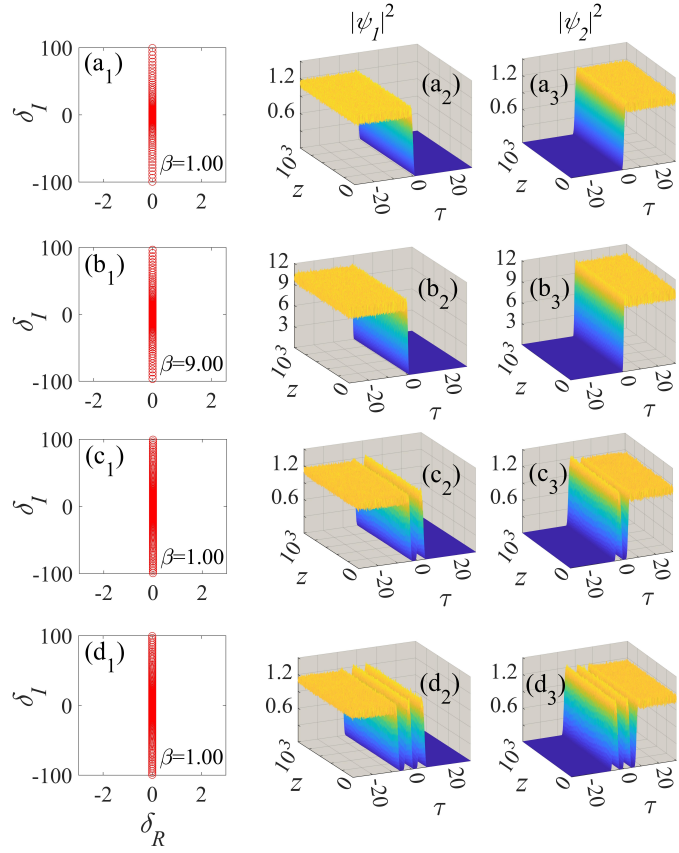}
\caption{Linear-stability spectra and perturbed evolution of the single- and multi-kink PQ-DW solitons. The linear-stability spectra are plotted in panels (a$_{1}$)-(d$_{1}$), while the middle (a$_{2}$)-(d$_{2}$) and right-hand (a$_{3}$)-(d$_{3}$) columns show the stable evolution of the $\protect\psi _{1}$ and $\protect\psi _{2}$ components with initial random perturbations at a $10\%$ amplitude level.}
\label{fig3}
\end{figure}

To test the stability of the PQ-DW solitons, we computed eigenvalues and eigenmodes of small perturbations around the PQ-DW solitons, using the Newton-conjugate gradient method, and simulated their perturbed evolution. To this end, perturbed solutions were introduced as
\begin{equation}
\Psi _{1}=e^{-i\beta z}\left[ \psi _{1}\left( \tau \right) +u_{1}\left( \tau\right) e^{\delta z}+u_{2}^{\ast }\left( \tau \right) e^{\delta ^{\ast }z}\right],
\label{Perturbation1}
\end{equation}
\begin{equation}
\Psi _{2}=e^{-i\beta z}\left[ \psi _{2}\left( \tau \right) +v_{1}\left( \tau\right) e^{\delta z}+v_{2}^{\ast }\left( \tau \right) e^{\delta ^{\ast }z}\right],
\label{Perturbation2}
\end{equation}
where $\psi _{1,2}$ represents the unperturbed soliton, defined as per Eq. (\ref{Psi1Psi2}), while $u_{1,2}$ and $v_{1,2}$ are small perturbations with the respective complex eigenvalue $\delta $ and $\ast $ standing for the complex conjugate. The substitution of expressions (\ref{Perturbation1}) and (\ref{Perturbation2}) in Eqs. (\ref{NLSE3}) and (\ref{NLSE4}) and linearization leads to the system of coupled equations:

\begin{equation}
i\delta u_{1}=+Lu_{1}+\psi _{1}^{2}u_{2}+2\psi _{1}\psi _{2}\left( v_{1}+v_{2}\right),
\label{Linearization_a}
\end{equation}

\begin{equation}
i\delta u_{2}=-\psi _{1}^{2}u_{1}-Lu_{2}-2\psi _{1}\psi_{2}\left( v_{1}+v_{2}\right),
\label{Linearization_b}
\end{equation}

\begin{equation}
i\delta v_{1}=+2\psi _{1}\psi _{2}\left( u_{1}+u_{2}\right)+Lv_{1}+\psi _{2}^{2}v_{2},
\label{Linearization_c}
\end{equation}

\begin{equation}
i\delta v_{2}=-2\psi _{1}\psi _{2}\left( u_{1}+u_{2}\right) -\psi _{2}^{2}v_{1}-Lv_{2},
\label{Linearization_d}
\end{equation}
\begin{equation}
L\equiv -\beta +\frac{1}{24}\frac{d^{4}}{d\tau ^{4}}+2\left( \psi_{1} ^{2}+ \psi _{2} ^{2}\right).
\label{Linearization_L}
\end{equation}

Equations (\ref{Linearization_a})-(\ref{Linearization_d}) were solved by means of the Fourier collocation method \cite{yang2010nonlinear}. The solitons are unstable if there are eigenvalues with $\mathrm{Re}(\delta )>0$. The results demonstrate that the PQ-DW solitons with the single- and multi-kink structure are completely stable, see examples of the linear-stability spectra in Figs. \ref{fig3}(a$_{1}$)-\ref{fig3}(d$_{1}$). At variance with classical bright scalar solitons, the topological nature of DW makes them strongly robust with respect to external perturbations such as temporal or amplitude fluctuations \cite{gilles2017polarization}.

\begin{figure}[th]
\centering
\includegraphics[width=0.5\linewidth]{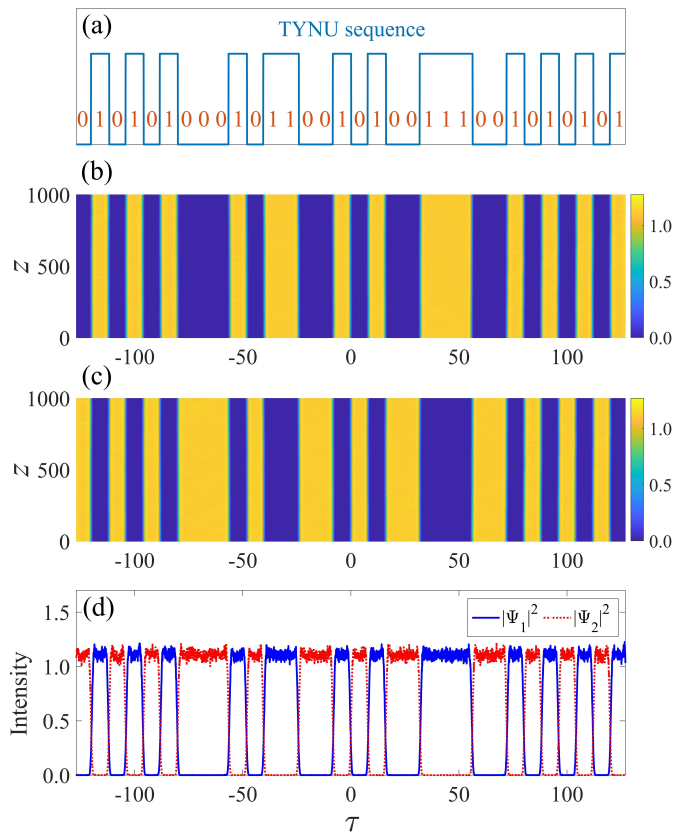}
\caption{The potential use of PQ-DW arrays for encoding data streams in optical telecommunications. (a) A schematic representation of the PQ-DW coding of a $32$- bit sequence. (b,c) The stable evolution of the $\Psi _{1}$ and $\Psi _{2}$ components with $\protect\beta =1$ under the action of random perturbations at the $10\%$ amplitude level. (f) Intensity output profiles corresponding to (b) and (c).}
\label{fig4}
\end{figure}

Direct stability tests were performed by simulations of Eqs. (\ref{NLSE3}) and (\ref{NLSE4}) with input taken as per Eqs. (\ref{Perturbation1}) and (\ref{Perturbation2}) at $z=0$, using the Runge-Kutta method. Generic examples of the stable evolution of the single- and multi-kink PQ-DW solitons are displayed in Figs. \ref{fig3}(a$_{2}$)-\ref{fig3}(b$_{2}$) and \ref{fig3}(a$_{3}$)-\ref{fig3}(b$_{3}$). The results confirm that they remain stable at least up to $z=1000$ under the action of random initial perturbations, even with a relatively large amplitude of $10\%$. The stable propagation of the kink-antikink-kink and antikink-kink-antikink PQ-DW solitons of Fig. \ref{fig1}(b) has been numerically confirmed up to $z=1000$, as shown in Figs. \ref{fig3}(c$_{2}$) and \ref{fig3}(c$_{3}$). Figs. \ref{fig3}(d$_{2}$) and \ref{fig3}(d$_{3}$) corroborate the robust propagation of the stable multi-kink PQ-DW solitons of Fig. \ref{fig1}(c).

Finally, we assessed the possibility of encoding data into arrays of the PQ-DW solitons, that may be used for optical telecommunications, as proposed in Refs. \cite{haelterman, gilles2017polarization}. To illustrate the principle, a 32 bit ASCII sequence encoding the acronym of Taiyuan Normal University (TYNU) is shown in Fig. \ref{fig4}(a). It is
encoded into the chain of PQ-DW solitons, which then propagates. Due to the robust balance between the PQ GVD, SPM and XPM, the temporal profile of the soliton chain remains stable, at least, up to $z=1000$ under the action of random perturbations at the $10\%$ amplitude level, as shown in Figs. \ref{fig4}(b,c) and confirmed by the output profiles in Fig. \ref{fig4}(d). Additional results demonstrate that reducing separation $\Delta \tau $ between adjacent DWs in the array enhances the interaction between them, eventually leading to instability at $\Delta \tau <4$.

\section{Conclusion}
In conclusion, we have reported the existence of the new type of DW (domain wall) solitons in optical media combining PQ (pure-quartic) GVD with the SPM and XPM nonlinearities. They exist in single- and multi-kink forms, which are completely stable against perturbations. We have also demonstrated that arrays of single-kink PQ-DW solitons can be used for the design of robust data-encoding schemes in optical telecommunications, which are stable against amplitude and temporal jitter. The present findings suggest straightforward experimental realization of individual and arrayed DWs. This is quite feasible, given recent advancements in intra- and extra-cavity spectral-temporal nonlinear shaping methods \cite{runge2020pure,han2024pure,https://doi.org/10.1002/lpor.202401714}. Note that the current model neglects odd-order dispersion terms (e.g., the third-order dispersion) and fiber loss -- in particular, because the loss can be compensated by distributed amplification. Preliminary results (to be reported elsewhere) suggest that the DWs and DW arrays remain essentially stable under the action of these effects, provided that they are not too strong. Thus, our findings expand the understanding of temporal PQ solitons in optics, and suggest potential applications to optical data transmission and processing.

\begin{acknowledgments}
This research was supported by the National Natural Science Foundation of China (11805141); Shanxi Province Basic Research Program (202203021222250, 202303021211185); Israel Science Foundation (grant No. 1695/22).
\end{acknowledgments}

\end{document}